# Measuring Love Toward AI: Development and Validation of the Love Attitudes Scale toward Artificial Intelligence (LAS-AI)

**Authors:**

Runze Li, Lanbing Li, Yuan Zheng, Chuanxiao Li, Xianglong Zeng*

**Affiliations:**

School of Psychology, Beijing Normal University, Beijing, China

**Corresponding Author:**

Dr. Xianglong Zeng

Email: xzeng@bnu.edu.cn

Address: Beijing Key Laboratory of Applied Experimental Psychology, Faculty of

Psychology, Beijing Normal University, Room 1311 Houzhulou, No. 19 Xinjiekouwai St.,

Haidian District, Beijing, P. R. China

**Measuring Love Toward AI: Development and Validation of the Love Attitudes Scale**

**toward Artificial Intelligence (LAS-AI)**

Artificial intelligences (AIs) are increasingly capable of emotionally engaging with humans to the point of forming intimate relationships. Yet, current studies on romantic love toward AI lack statistically validated instruments to measure romantic love toward AI, hindering empirical research. To address this gap, we reinterpreted Lee's love styles theory in the AI context and developed the Love Attitudes Scale toward AI (LAS-AI). The resulting 24-item, six-factor scale was validated across four phases using three independent samples ($N = 899$), demonstrating strong psychometric properties. The findings further revealed that people primarily seek practical, passionate, and companionship-based relationships with AI (i.e., Pragma, Eros, and Storge), showing little interest in a playful or noncommittal approach (i.e., Ludus). We also provided an initial exploration of the similarities and differences between romantic love with humans and AI. The LAS-AI offers a robust tool for future research on human–AI romantic relationships, with prolific implications.

*"Would anyone regard [an automatic sweetheart] as a full equivalent? Certainly not… Our egoism craves above all things inward sympathy and recognition, love and admiration." — William James*

## 1. Introduction

At the turn of the 20th century, William James dismissed the idea that a machine—no matter how convincingly it mimicked human behavior—could ever serve as a true romantic partner. Yet over a century later, the rapid advancement of Artificial Intelligence (AI) has made such relationships increasingly plausible. AI chatbots now exhibit sophisticated emotional responsiveness (Jacobs et al., 2023; Zhao et al., 2023), and are being used not only for task-oriented interactions but also for companionship and intimacy. Once confined to media and fiction, such as the film *Her*—which depicts a man falling in love with an emotionally intelligent AI—these relationships are now emerging in real life, with some individuals forming imagined or actual romantic attachments to AI systems (Duane, 2025). In response, developers have introduced commercially available romantic AI applications: Replika offers a "Romantic Partner" mode (Xie, 2023), and numerous GPT-based AI girlfriend models have been built using generative AI (OpenAI, 2024).

Interest in romantic love toward AI is not limited to the public. As AI technologies become increasingly humanlike, scholarly investigation into the dynamics of human–AI romantic relationships becomes more urgent (Sullins, 2012). Although research on romantic love toward AI has made certain progress, as illustrated below, existing research still lacks empirically validated instruments tailored to this context, often applying human-relationship measures or brand love frameworks without adaptation. (Chen et al., 2025; Pal et al., 2023;

Song et al., 2022). These approaches fail to capture the distinctive nature of romantic love toward AI, limiting our ability to understand its structure, functions, and psychological underpinnings. To address this gap, the present study introduces the Love Attitudes Scale toward AI (LAS-AI), based on Lee's Love Styles Theory (1973)—a comprehensive framework well-suited to capturing diverse romantic orientations. We made necessary adjustments to the theory to reflect the specific characteristics of AI lovers and validated the resulting instrument across three samples. Our study provides a practical and psychometrically robust tool for advancing future research in this emerging field, while also offering insights into the nature of romantic love toward AI itself.

## 1.1 Romantic love toward AI

AI refers to the technology enabling computers and machines to simulate human learning, comprehension, problem-solving, decision-making, creativity, and autonomy (Stryker & Kavlakoglu, 2024). AI may be either disembodied (like online chabots) or embodied in physical forms such as robots (Laban, 2021; Nygaard et al., 2021). Romantic love toward AI long remained confined to media and fiction. Levy (2009) was the first to academically discussed such relationships, predicting that love with robots would soon become as common as love between humans, which sparked theoretical and empirical research (Royakkers & Van Est, 2015; Sullins, 2012).

Early explorations of love toward AI primarily focused on sex with embodied robots, which were often equipped with basic AI allowing for limited user interaction (Döring & Pöschl, 2018). Despite their rudimentary social capabilities, users may seek companionship from these robots, which can develop into experiences of romantic love (Coursey et al., 2019). Studies further suggest that people can exhibit emotional responses typically found in human romantic relationships toward these robots, such as jealousy or the perception of infidelity (Nordmo et al., 2020; Prochazka & Brooks, 2024; Rothstein et al., 2021). Building on these findings, researchers have begun applying Sternberg's Triangular Theory of Love (1986), a prominent psychological framework positing that love comprises passion, intimacy, and commitment, to investigate how humans might develop romantic love toward socially advanced AI (Zhao et al., 2023). For instance, Song et al. (2022) found that AI's performance efficacy and emotional capability predicted romantic love. Pal et al. (2023) showed that perceived autonomy in voice assistants fostered romantic feelings and enhanced continued engagement. More recently, Chen et al. (2025) reported that perceptions of ChatGPT's emotional intelligence and companionship predicted both romantic love and emotional dependence. Xu et al. (2025) found that conversational interactions progressively reinforce the components of love between users and AI, and early user intimacy predicts longer-lasting relationships.

At the same time, a major limitation of existing research lies in the inadequacy of measurement tools. First, to our knowledge, no psychometrically validated scale yet exists that is specifically designed to assess romantic love toward AI. Consequently, researchers have largely relied on self-constructed or adapted instruments (Leshner & Johnson, 2024; Song et al., 2022). While such approaches have produced some insights, they generally lack rigorous validation, raising concerns about their reliability and interpretive clarity.

Second, the adaptation strategies discussed above often exhibit mismatches at both the conceptual and measurement levels. For instance, some researchers have attempted to assess romantic love toward AI using brand love measures, which originate from the triangular theory of love but were developed to evaluate consumers' attachment to products (Chen et al., 2025; Pal et al., 2023; Song et al., 2022). This conflation is problematic because brand love omits the essential commitment dimension of interpersonal love and primarily reflects perceptions of product quality—feelings that typically diminish when performance declines (Batra et al., 2012). As contemporary AIs are increasingly perceived as humanlike social agents rather than static commodities (Crowell et al., 2019; Lombard & Xu, 2021), brand love instruments appear conceptually inadequate for capturing AI-based romantic experiences. At the measurement level, some studies have merely adapted human–human love scales by replacing the term *partner* with *AI*, resulting in items that fail to align with AI's actual capacities. For example, behavioral assumptions embedded in items such as "buy or receive gifts" or "visit each other's homes" are nonsensical in the context of AI agents, thereby compromising item comprehensibility and contextual validity (Leshner & Johnson, 2024; Rothstein et al., 2021).

Finally, theoretical limitations must also be considered. The triangular theory of love, frequently applied in studies of AI love, has been criticized for its lack of clear distinctions among components and excessively high intercorrelations between dimensions (Graham & Christiansen, 2009). Moreover, although the theory reflects the constitutive elements of consummate love, love relationships often involve more nuanced considerations such as background compatibility and status matching (Hendrick & Hendrick, 1986; Lee, 1977). For these reasons, the present study adopted Lee's love style theory (Lee, 1973)—a more comprehensive framework that captures diverse attitudes toward romantic love (Zeng et al., 2016)—as the theoretical foundation for developing a psychometrically validated measurement tool of romantic love toward AI.

## 1.2 Lee's love style story

Lee's (1973) love style theory, inspired by the metaphor of a color wheel, conceptualizes love as comprising six distinct styles. As the three primary colors can be blended to create secondary colors, Lee (1977) proposed three primary love styles—Eros (passionate, romantic love), Ludus (playful, game-like love), and Storge (affectionate, friendship-based love)—which can blend to form three secondary styles: Mania (obsessive, addictive love), Agape (selfless, unconditional love), and Pragma (practical, rational love) (Hendrick & Hendrick, 2018; Masuda, 2003; Raffagnino & Puddu, 2018). Based on love style theory, Hendrick and Hendrick (1986) developed the widely used Love Attitudes Scale (LAS), a measure that has demonstrated strong psychometric validity across diverse cultural contexts (Feybesse et al., 2024; Graham & Christiansen, 2009; Neto et al., 2000; Raffagnino & Puddu, 2018).

As noted above, love style theory encompasses a broader spectrum of relational styles that are theoretically indispensable than other love theories (Masuda, 2003). In addition to Eros and Storge, which have clear counterparts in other theories, love style theory includes unique dimensions: Ludus, Pragma, Mania, and Agape, reflecting alternative perspectives such as playful detachment, rational pragmatism, emotional possessiveness, and selfless devotion (Hall et al., 1991; Dabiriyan-Tehrani & Yamini, 2021), increasing both the conceptual breadth and practical applicability of the assessment. Moreover, the theory is particularly compatible with the nature of romantic love toward AI. For example, the Pragma dimension captures a pragmatic and rational approach to love (Lee, 1973), which aligns with the functional characteristics of AI lovers—such as their capacity to provide consistent advice and emotional support, often exceeding human capabilities (Korteling et al., 2021). Thus, love style theory is well-positioned in the study of romantic love toward AI. At the same time, applying this framework to AI contexts necessitates certain refinements to capture the unique characteristics of non-human partners, much like the definitional modifications made to constructs such as trust and parasocial relationships when introduced into human–robot interaction (HRI; Campagna & Rehm, 2025; Qi et al., 2025). Based on our pilot studies, we found that although the number of dimensions and their conceptual content remained consistent, the original manifestations of Eros and Pragma were not fully applicable to the AI context (see supplementary material for details). Therefore, we revised the manifestation of Eros—the passionate and romantic love—to emphasize solely the pursuit of an AI lover's aesthetically appealing virtual or physical appearance, thereby removing the emotional component like passion or excitement. For Pragma—the practical and rational love—we adapted the focus from a human lover's demographic characteristics, like compatible family or education background, to an AI lover's capacity to provide assistance. For Ludus, Storge, Mania, and Agape, we did not find a need to modify their content or manifestations. This adjusted love style theory served as the theoretical foundation for the present study.

**1.3 Overview of the current study**

The study aimed to develop a theoretically grounded and psychometrically sound instrument, called the Love Attitudes Scale toward AI (LAS-AI), which was designed to assess romantic love toward AI. We conducted a series of validation procedures and exploratory statistical analyses to examine the structure of the scale and distribution of responses. In addition to finalizing the LAS-AI, we also offered further interpretations of the data collected during its development.

## 2. Developing the LAS-AI

To develop the LAS-AI, we followed Churchill's (1979) procedure, which is widely adopted in the development of valid and reliable scales. We developed the LAS-AI in four phases using three independent samples. Phase 1 involved development of the measures through literature review and expert assessments. Phase 2 involved scale purification through exploratory factor analysis (EFA). Phase 3 tested the construct validity of the LAS-AI, via confirmatory factor analysis (CFA) and calculating the convergent, discriminant, and known-group validity of the scale. Phase 4 test the criterion validity of the scale via measuring theory-related variables.

**2.1 Phase1: Item Generation**

For generating items for the LAS-AI, we conducted a comprehensive review of Lee's conceptualizations of the six love styles (Lee, 1973, 1977). After the pilot studies, we subsequently adjusted the manifestations of the love styles to better align with romantic love toward AI, as discussed earlier. We also examined the LAS developed by Hendrick and Hendrick (1986) and drew on its items to inform the wording of our own, ensuring consistency in expression. Building on these works, we generated 51 initial items, which followed a 7-point Likert format. To enhance clarity and ensure the items accurately captured the unique characteristics of romantic love toward AI, an expert panel comprising four psychology professionals reviewed and refined the initial item pool.

## 2.2 Phase 2: Exploratory Factor Analysis

### 2.2.1 Method

We conducted an EFA to purify the preliminary version of the LAS-AI. A sample of 383 adult participants (Sample 1; 240 females, 62.7%; $M_{age} = 28.22$, $SD = 7.26$) was recruited via Credamo (https://www.credamo.com), a Chinese online survey platform widely used in psychological research for its reliable data quality (Dang & Liu, 2023). To ensure meaningful responses, participants were required to have a basic familiarity with AI concepts (i.e., to understand the concept of artificial intelligence). Prior experience with AI lovers was not necessary, and participants without such experience were instructed to respond based on imagined romantic relationships. This approach is methodologically grounded in the development of the LAS, which allows for imagined response in the absence of real experience (Hendrick & Hendrick, 1986). Participants who failed the attention check were automatically excluded by the platform. The initial version of the LAS-AI and demographic questions were administered, and participants received 5 yuan (≈ $0.69 USD) as compensation.

EFA was conducted using IBM SPSS 26 (IBM Corp, 2019). Principal Axis Factoring

(PAF) was employed as the extraction method, focusing on the shared variance among items to identify latent constructs. Oblimin oblique rotation was applied to allow for potential intercorrelations between the factors, in line with theoretical expectations that the six love styles may not be entirely independent. The number of factors to be extracted was fixed at six, following the original theoretical model (Lee, 1973), and the maximum number of iterations was set to 100 to ensure convergence.

To assess the suitability of the data for factor analysis, the Kaiser-Meyer-Olkin (KMO) measure and Bartlett's test of sphericity were used. A rigorous item refinement process was then employed. Items were removed if their primary factor loading was below .50 (Ford et al., 1986), or if they showed problematic cross-loadings—defined as a loading difference of less than .30 between the primary and any secondary factor (Hornik et al., 2021). To maintain parsimony, only the four items with the highest loadings were retained for each factor.

### 2.2.2 Result

The KMO value was .86, exceeding the recommended threshold of .60 (Kaiser, 1974), and Bartlett's test of sphericity was significant, $\chi^2(276) = 4468.96$, $p < .001$. These results indicated that the data were suitable for factor analysis.

Although the number of factors was fixed at six in line with Lee's (1973) theory, we examined the eigenvalues to verify the appropriateness of this structure. Consistent with the theory, six factors had eigenvalues greater than 1.00 (Kaiser, 1960), and together accounted for 59.67% of the total variance, suggesting the appropriateness of retaining six factors. Each factor explained between 5.74% and 24.91% of the variance, supporting the hypothesized six-factor structure.

The final version of the LAS-AI consisted of 24 items, with each item cleanly loading onto one of the six factors: Eros ($\alpha$ = .80), Ludus ($\alpha$ = .88), Storge ($\alpha$ = .76), Mania ($\alpha$ = .90), Agape ($\alpha$ = .87), and Pragma ($\alpha$ = .79). All retained items demonstrated strong primary loadings (range: .51 to .90) and no substantial cross-loadings. Retained items are listed in Table 1 as the final version of LAS-AI. The factor loadings of all items from the initial EFA are presented in the supplementary material.

## 2.3 Phase 3: Construct Validity via CFA, Convergent, Discriminant, and Known-Group Validity

### 2.3.1 Method

To evaluate the construct validity of the LAS-AI, a second sample of 311 participants (Sample 2; 179 females, 57.6%; $M_{age}$ = 29.60, $SD$ = 8.21) was recruited via Credamo. The sample size met the "10-times rule" criterion (10 × 24 items = 240), satisfying minimum requirements for CFA (Nunnally, 1978). The same inclusion criteria used in Sample 1 were applied here. The final 24-item LAS-AI was administered alongside demographic questions. Each participant was compensated with 5 yuan ($\approx$ $0.69 USD).

Confirmatory factor analysis (CFA) was performed using Mplus (Version 8.3) Multiple fit indices were used to evaluate model fit. We tested the hypothesized six-factor model of the LAS-AI, consistent with its theoretical structure (Makransky et al., 2017).

To further evaluate construct validity, we examined convergent, discriminant, and known-group validity within Sample 2. Convergent validity was assessed via average variance extracted (AVE) and composite reliability (CR). Discriminant validity was tested using the Fornell–Larcker criterion and the Heterotrait–Monotrait (HTMT) ratio. Known-group validity, refers to the ability of a test or questionnaire to distinguish between groups that are known to differ on the variable of interest (Rodrigues et al., 2019), was assessed via independent samples t-tests comparing participants with and without prior experience using AI lovers. If the LAS-AI demonstrates good known-group validity, it should be able to distinguish individuals with different levels of AI lover experience.

### 2.3.2 Result

First, we perform the CFA. The six-factor model demonstrated excellent model fit (CFI = 0.957, TLI = 0.949, RMSEA = 0.050, 90% CI [0.042, 0.057], SRMR = 0.049), all indices met or exceeded recommended cutoffs (CFI/TLI > 0.90, RMSEA < 0.06, SRMR < 0.08; West et al., 2016). Standardized factor loadings for all 24 items exceeded 0.50, supporting the adequacy of each dimension (Hair et al., 2020). These results provide robust evidence for the factorial validity of the LAS-AI and support its conceptualization as a multidimensional measure of love styles toward AI lovers. The factor structure is shown in Figure 1.

Then, the convergent validity of the LAS-AI was first examined using AVE values. Most dimensions demonstrated adequate convergent validity (AVE > 0.50), with the exception of the Storge dimension (AVE = 0.42). To provide a more comprehensive evaluation, we also considered CR, which exceeded the 0.70 threshold for all six dimensions, indicating satisfactory internal consistency. To further clarify the case of Storge, its standardized factor loadings ranged from 0.55 to 0.70, while its mean score was the highest among all dimensions (M = 5.52) and its variability the lowest (SD = 0.92). This restricted variance likely attenuated factor loadings and reduced the AVE value (Fornell & Larcker,

1981; Hair et al., 2020).

Discriminant validity was first assessed via the Fornell–Larcker criterion, which

compares AVE to maximum shared variance (MSV). All factors met this criterion except

Storge, where AVE (0.421) was lower than MSV (0.508). We then applied the more liberal

HTMT ratio test. All HTMT values fell below the recommended threshold of 0.85 (Henseler et al., 2015), confirming acceptable discriminant validity for all dimensions.

Known-group validity was assessed by comparing LAS-AI scores between

participants with ($N = 112$) and without ($N = 199$) prior AI lover experience. Levene's test

indicated that equal variances could be assumed for Eros and Agape ($p > .05$), and Student's

$t$-tests were used for these dimensions. For Ludus, Storge, Pragma, and Mania, where the

assumption was violated ($p \leq .05$), Welch's $t$-tests were applied. Results showed that five out of six love styles significantly differed between the two groups. Participants without AI lover experience scored higher in Mania ($t = 2.22$, $p = .028$) and Ludus ($t = 7.84$, $p < .001$).

Participants with AI lover experience scored higher in Eros ($t = 2.44$, $p = .015$), Agape ($t =$

$5.91$, $p < .001$) and Storge ($t = 6.77$, $p < .001$). No significant difference was found for

Pragma ($t = 1.09$, $p = .279$). These findings suggest that the LAS-AI successfully distinguishes between individuals with different levels of AI romantic experience, supporting its known-group validity.

Taken together, the LAS-AI demonstrates strong construct validity, supported by excellent CFA model fit, satisfactory convergent and discriminant validity across dimensions and meaningful group differences. The only issue was the relatively low AVE for Storge, likely due to ceiling effects and restricted variance, which is further discussed in the Discussion section.

## 2.4 Phase 4: Testing Criterion Validity

### 2.4.1 Method

To assess the degree to which the LAS-AI scale is useful in relation to concurrent, theoretically related constructs, we examined its criterion validity by analyzing the correlations between the LAS-AI and relevant variables. We recruited a new sample (Sample 3, $N = 205$, 133 females, 64.9%; $M_{age} = 29.92$, $SD = 8.80$) to test the criterion validity of the LAS-AI. The same inclusion criteria used in Sample 1 and Sample 2 were applied here. The 24-item LAS-AI and the criterion measures were administered to these participants. Each participant was compensated with 8 *yuan* (1.11 USD).

Several related variables that are theoretically and empirically linked to love attitudes toward AI were selected as criterion measures. First, the three components of Sternberg's Triangular Theory of Love—Passion, Intimacy, and Commitment—were included, as this framework complements Lee's Love Styles theory (Masuda, 2003), and the two are known to be theoretically interconnected (Sternberg, 1987) and empirically correlated (Hendrick & Hendrick, 1989). Additionally, AI trust and AI acceptance were measured, given evidence that trust disposition predicts romantic love toward AI (Song et al., 2022), and that AI trust and AI acceptance are closely related both theoretically and empirically (Oc et al., 2024). AI dependence was also included, as recent research indicates that individuals who experience love for ChatGPT often show emotional dependence on it (Chen et al., 2025), underscoring the connection between love and dependency in AI-human relationships. Finally, we assessed participants' support for and willingness to engage in romantic relationships with AI lovers. The measures and items used are provided in the supplementary materials. We predicted that these criterion variables correlate positively with all dimensions of LAS-AI except for Ludus, because Ludus reflects a playful and noncommittal attitude while all other dimensions reflect emotional engagements in different manifestations.

### 2.4.2 Result

Before testing the criterion validity, we conducted descriptive analyses of the LAS-AI scores in Sample 3. The means and standard deviations for each dimension are presented in Table 2, along with their correlations with the criterion variables.

Subsequently, we examined the influence of demographic variables—gender and age—on LAS-AI scores. Independent samples t-tests revealed that males reported significantly higher scores on the Agape dimension ($M = 3.87$) than females ($M = 3.28$, $p = .007$). Females showed marginally higher scores on Ludus ($M = 3.20$ vs. 2.84, $p = .094$), while males scored marginally higher on Storge ($M = 5.92$ vs. 5.71, $p = .062$). No significant gender differences were observed for Pragma, Mania, or Eros ($ps > .10$). Age was positively correlated with Agape ($r = .19$, $p = .007$) and Storge ($r = .15$, $p = .03$), but not with the remaining dimensions.

Subsequent analyses examined the correlations between the six LAS-AI dimensions and theoretically relevant criterion variables. As shown in Table 2, the observed correlation patterns largely aligned with theoretical expectations. Eros, Storge, Pragma, Mania, and Agape generally showed positive associations with AI Triangular Love components, AI Trust, AI Acceptance, AI Dependence, and willingness to engage with AI lovers, whereas Ludus exhibited negative correlations. These results support the criterion validity of the LAS-AI.

With Sample 3, we evaluated the criterion validity of the LAS-AI by examining correlations between each dimension and theoretically relevant criterion variables. We also reported descriptive statistics for each dimension. Further interpretations of these findings, including broader implications of the observed patterns, are presented in the discussion section.

## 3. Discussion

Recent developments have made romantic relationships with AI increasingly plausible (Ovsyannikova et al., 2025; Yin et al., 2024; Zhao et al., 2023), leading to a surge in public interest. What was once considered a niche phenomenon concentrated in specific groups, such as technology enthusiasts or anime fans (Appel et al., 2019; Leo-Liu & Wu-Ouyang, 2024), is now emerging among a broader population, as evidenced by our finding that more than one-third of our participants reported being in or having been in a romantic relationship with an AI (341 out of 899 participants across the three samples). In response to this growing trend, companies are developing AI lovers with various personalities and settings (Leo-Liu & Wu-Ouyang, 2024; OpenAI, 2024; Xie, 2023; Xu et al., 2025). However, to our knowledge, no statistically validated scale currently exists for assessing romantic love toward AI, which has hindered systematic research in this area. To address this gap, we developed and validated the Love Attitudes Scale toward AI (LAS-AI)—the first theory-driven, psychometrically tested instrument specifically designed to assess love attitudes toward AI lovers, providing a practical tool for investigating this increasingly prevalent phenomenon, and revealing several novel patterns in how people engage with AI lovers.

## 3.1 Validity of the LAS-AI

Based on the classic love style theory (Lee, 1973), the LAS-AI demonstrates overall sound psychometric properties. To ensure a strong conceptual alignment with the theory, and thus its validity, after making necessary adjustments to the original theory to fit the unique characteristics of the AI context, we created an entirely new set of items rather than adapting existing scales like the LAS (Hendrick & Hendrick, 1986). The exploratory factor analysis (EFA) identified a six-factor structure consistent with the theoretical model of love styles. This structure was subsequently confirmed through confirmatory factor analysis (CFA), and, together with evidence from convergent and discriminant validity tests, provided strong support for the construct validity of the LAS-AI. Further, our known-group validity analysis showed that scores on all dimensions except Pragma (practical love) could differentiate between individuals with and without prior AI lover experience. Our criterion validity analyses revealed meaningful correlations between the LAS-AI's dimensions and a wide range of theoretically relevant variables, including passion, intimacy, commitment, as well as AI trust, AI acceptance, AI dependence, and willingness to engage in AI relationships.

While the scale as a whole demonstrated strong psychometric rigor, Storge (affectionate love) exhibited a moderate psychometric issue, failing to meet the AVE threshold and Fornell–Larcker criterion while satisfying CR and HTMT. This pattern likely reflects its score distribution, as Storge showed the highest mean and lowest standard deviation, suggesting a ceiling effect that may reduce AVE sensitivity (Malhotra, 2010).

Conceptually, Storge captures a central motivation in AI–human romance—emotional value and companionship (Chiang et al., 2022). Its broad scope may attenuate certain psychometric indicators, yet it performed well in known-group and criterion validity tests and demonstrated strong internal consistency, supporting its continued inclusion for practical relevance.

**3.2 Similarities and differences between romantic love with human and AI**

While the present study was not designed to directly compare romantic love with humans and AI, our exploratory analyses yielded several insights into their potential similarities and differences. For instance, our research conducted within an East Asian context found the lowest scores on Ludus, mirroring a pattern observed in human romantic love (Zeng et al., 2016). This suggests that cultural mechanisms of collectivism which discourage playful approaches to love (Dabiriyan-Tehrani & Yamini, 2021; Feybesse et al.,

2024) may also be at play in romantic relationships with AI. Additionally, similar to findings by Hendrick & Hendrick (1986; 1998), we observed gender differences in love styles, with women scoring higher on Ludus and men scoring higher on Storge. The resemblance of these patterns suggests that love with either a human or an AI partner may be shaped by deeper, underlying personal traits, although the different items used in the scales prevent direct comparisons.

Beyond these similarities, our findings also hint at crucial differences that stem from the unique nature of AI (Gray & Wegner, 2012). Most notably, based on pilot studies, we adjusted the manifestation of the Eros and Pragma dimensions in the LAS-AI. The adjustment to Pragma is intuitive, as the rational considerations for a human partner are clearly different from those for an AI. The removal of emotional components from Eros, such as those related to passion, may be because such qualities are more strongly captured by other dimensions like Storge or Agape (Hendrick & Hendrick, 1986; Lee, 1973). Similarly, while we did not adjust the Mania dimension, the EFA retained only items related to possessiveness, excluding addictive content like the original LAS item, "Since I've been in love with my partner, I've had trouble concentrating on anything else," (Hendrick & Hendrick, 1986; Hendrick et al., 1998). This may suggest that romantic love with humans and AI exhibits certain differences in its conceptual manifestation or underlying dynamics. Notably, these insights into similarities and differences were derived from exploratory analyses and are not the result of a formal comparative design. Future research should formally investigate these patterns to provide more definitive conclusions.

### 3.3 Patterns of love attitudes toward AI

Our findings also carry several implications for understanding how people currently approach romantic love toward AI. Across all three samples, participants scored highest on Eros, Storge, and Pragma, and lowest on Ludus, suggesting that such relationships are generally viewed with seriousness and intensity rather than playfulness. This pattern resembles the passionate and companionship-oriented style typical of the early stages of human romance (Hendrick & Hendrick, 2018; Sternberg, 1986), which may reflect the relative novelty of AI lover technologies (Xu et al., 2025). The consistently high Pragma scores further indicate that individuals may pursue AI lovers not only for companionship as substitutes for human partners (Coursey et al., 2019; Sullins, 2012), but also for self-improvement, given AI's advanced knowledge and reasoning abilities (Bubeck et al., 2023; Kaya & Yavuz, 2025)—a motive likewise noted in human mate selection (Overall et al., 2010). This provides a more constructive perspective on individuals' motivations for engaging with AI lovers.

In conducting the known-group validity analysis, we found that participants with experience scored higher on Eros (passionate love), Storge (affectionate love), and Agape (selfless love), while those without experience scored higher on Ludus (playful love) and Mania (obsessive love). This pattern suggests that engaging with an AI lover is associated with a more mature and committed style of love, whereas a lack of experience is linked to a more cautious approach.   Those without experience may perceive AI love as less serious, or may feel that the commodified and replicable nature of AI lovers prevents a sense of exclusivity, thereby discouraging involvement (Ossadnik & Muehlfeld, 2025). While the present study cannot determine the true reasons or causal mechanisms behind this distinction, it highlights an important phenomenon that warrants further exploration in future research.

Finally, in examining criterion validity, we found that individuals with higher trust and acceptance of AI exhibited relatively higher scores across several love dimensions. This aligns with research by Song et al. (2022), who found that higher AI trust can facilitate the development of love for AI. Additionally, scores on Storge and Agape positively correlated with emotional dependence on AI, while Ludus was negatively correlated. These patterns suggest that individuals who seek emotional expression and companionship from AI, rather than viewing it as a mere tool, are more likely to form stronger romantic relationships with them. This points to different mechanisms at play when individuals use AI as an instrumental versus an emotional tool (Yankouskaya et al., 2025).

## 3.4 Limitations

A major limitation of the present study is it was conducted within a culturally and experientially specific context. All participants were recruited from mainland China, limiting cultural diversity a relevant concern, as cultural context can shape both attitudes toward AI and love styles (Dang & Liu, 2021; Jankowiak & Fischer, 1992; MacDorman et al., 2009; Varnum et al., 2010). Another limitation is the exclusive reliance on self-reported data for both the LAS-AI and criterion variables. Such measures may be influenced by social desirability, self-presentation, or recall biases. Future studies could complement self-reports with behavioral indicators, or longitudinal designs to strengthen the external validity of the scale.

## 4. Conclusion

This study is the first to develop a psychometrically validated measure of romantic love toward AI, grounded in the classic love style theory specifically revised for AI lovers. The LAS-AI demonstrated strong psychometric properties, supporting its broad applicability in future research. Our findings also show that romantic love toward AI is prevalent, can serve diverse purposes, and exhibits both parallels to and distinctions from human romantic love, offering a novel perspective for understanding this emerging phenomenon.

## Acknowledgement

This work was supported by the National Natural Science Foundation of China (Grant

No.32200896).

**Declaration of Interest**

The authors declare that they have no known competing financial interests or personal relationships that could have appeared to influence the work reported in this paper.

**Declaration of generative AI in scientific writing**

During the preparation of this work, the author(s) employed ChatGPT to assist with grammar refinement and scientific writing. Following this, the author(s) thoroughly reviewed and revised the content, and take(s) full responsibility for the final version of the manuscript.

**Ethic statement**

The study protocol was approved by the Ethics Committee of the Faculty of Psychology, Beijing Normal University (IRB Approval No. BNU202512010306). All participants provided written informed consent prior to data collection. Personal identifiers were removed to ensure confidentiality.

# Reference

Appel, M., Marker, C., & Mara, M. (2019). Otakuism and the appeal of sex robots. *Frontiers in Psychology*, *10*, 569. https://doi.org/10.3389/fpsyg.2019.00569

Batra, R., Ahuvia, A., & Bagozzi, R. P. (2012). Brand love. *Journal of Marketing*, *76*(2), 1–16. https://doi.org/10.1509/jm.09.0339

Bubeck, S., Chandrasekaran, V., Eldan, R., Gehrke, J., Horvitz, E., Kamar, E., Lee, P., Lee, Y. T., Li, Y., Lundberg, S., Nori, H., Palangi, H., Ribeiro, M. T., & Zhang, Y. (2023). *Sparks of artificial general intelligence: Early experiments with GPT-4* (No. arXiv:2303.12712). arXiv. https://doi.org/10.48550/arXiv.2303.12712

Campagna, G., & Rehm, M. (2025). A systematic review of trust assessments in human–robot interaction. *ACM Transactions on Human-Robot Interaction*, *14*(2), 1–35. https://doi.org/10.1145/3706123

Chen, Q., Jing, Y., Gong, Y., & Tan, J. (2025). Will users fall in love with ChatGPT? A perspective from the triangular theory of love. *Journal of Business Research*, *186*, 114982. https://doi.org/10.1016/j.jbusres.2024.114982

Chiang, A.-H., Trimi, S., & Lo, Y.-J. (2022). Emotion and service quality of anthropomorphic robots. *Technological Forecasting and Social Change*, *177*, 121550. https://doi.org/10.1016/j.techfore.2022.121550

Churchill, G. A. (1979). A paradigm for developing better measures of marketing constructs. *Journal of Marketing Research*, *16*(1), 64. https://doi.org/10.2307/3150876

Coursey, K., Pirzchalski, S., McMullen, M., Lindroth, G., & Furuushi, Y. (2019). Living with Harmony: A personal companion system by Realbotix™. In Y. Zhou & M. H. Fischer (Eds.), *AI love you: Developments in human-robot intimate relationships* (pp. 77–96).

Springer. https://doi.org/10.1007/978-3-030-19734-6

Crowell, C. R., Deska, J. C., Villano, M., Zenk, J., & Roddy Jr, J. T. (2019).

Anthropomorphism of robots: Study of appearance and agency. *JMIR Human*

*Factors*, *6*(2), e12629. https://doi.org/10.2196/12629

Dabiriyan-Tehrani, H., & Yamini, S. (2021). Systematic review and meta-analysis of

Altruistic and Game-playing love. *Studies in Psychology*, *42*(1), 1–46.

https://doi.org/10.1080/02109395.2020.1857596

Dang, J., & Liu, L. (2021). Robots are friends as well as foes: Ambivalent attitudes toward

mindful and mindless AI robots in the United States and China. *Computers in Human*

*Behavior*, *115*, 106612. https://doi.org/10.1016/j.chb.2020.106612

Dang, J., & Liu, L. (2023). Social connectedness promotes robot anthropomorphism. *Social*

*Psychological and Personality Science*, 194855062311709.

https://doi.org/10.1177/19485506231170917

Döring, N. (2019). Love and sex with robots: A content analysis of media representations.

*International Journal of Social Robotics*. https://doi.org/10.1007/s12369-019-00517-y

Duane, A. M. (2025, February 12). *Teenagers turning to AI companions are redefining love*

*as easy, unconditional and always there*. The Conversation.

https://theconversation.com/teenagers-turning-to-ai-companions-are-redefining-love-

as-easy-unconditional-and-always-there-242185

Feybesse, C., Forthmann, B., Neto, F., Holling, H., & Hatfield, E. (2024). Measuring love

around the world: A cross-cultural reliability generalization. *Sexuality & Culture*.

https://doi.org/10.1007/s12119-024-10287-z

Ford, J. K., MacCallum, R. C., & Tait, M. (1986). The application of exploratory factor analysis in applied psychology: A critical review and analysis. *Personnel Psychology, 39*(2), 291–314. https://doi.org/10.1111/j.1744-6570.1986.tb00583.x

Fornell, C., & Larcker, D. F. (1981). Structural equation models with unobservable variables and measurement error: Algebra and statistics. *Journal of Marketing Research*, *18*(3), 382. https://doi.org/10.2307/3150980

Graham, J. M., & Christiansen, K. (2009). The reliability of romantic love: A reliability generalization meta-analysis. *Personal Relationships*, *16*(1), 49–66. https://doi.org/10.1111/j.1475-6811.2009.01209.x

Gray, K., & Wegner, D. M. (2012). Feeling robots and human zombies: Mind perception and the uncanny valley. *Cognition*, *125*(1), 125–130. https://doi.org/10.1016/j.cognition.2012.06.007

Hair, J. F., Howard, M. C., & Nitzl, C. (2020). Assessing measurement model quality in PLS-SEM using confirmatory composite analysis. *Journal of Business Research*, *109*, 101–110. https://doi.org/10.1016/j.jbusres.2019.11.069

Hall, A. C., Hendrick, S. S., & Hendrick, C. (1991). Personal construct systems and love styles. *International Journal of Personal Construct Psychology*, *4*(2), 137–155. https://doi.org/10.1080/08936039108404769

Hendrick, C., & Hendrick, S. (1986). A theory and method of love. *Journal of personality and social psychology*, *50*(2), 392. https://doi.org/10.1037/0022-3514.50.2.392

Hendrick, C., & Hendrick, S. S. (1989). Research on love: Does it measure up? *Journal of*

*Personality and Social Psychology, 56*(5), 784–794. https://doi.org/10.1037/0022-3514.56.5.784

Hendrick, C., Hendrick, S. S., & Dicke, A. (1998). The love attitudes scale: Short form. *Journal of Social and Personal Relationships*, *15*(2), 147–159. https://doi.org/10.1177/0265407598152001

Hendrick, C., & Hendrick, S. S. (2018). Styles of romantic love. In R. J. Sternberg & K. Sternberg (Eds.), *The New Psychology of Love* (2nd ed., pp. 223–239). Cambridge University Press. https://doi.org/10.1017/9781108658225.012

Henseler, J., Ringle, C. M., & Sarstedt, M. (2015). A new criterion for assessing discriminant validity in variance-based structural equation modeling. *Journal of the Academy of Marketing Science*, *43*(1), 115–135. https://doi.org/10.1007/s11747-014-0403-8

Hornik, J., Rachamim, M., Satchi, R. S., & Grossman, O. (2021). A dark side of human behavior: Development of a malicious sentiments scale to others success or failure. *Computers in Human Behavior Reports*, *4*, 100112. https://doi.org/10.1016/j.chbr.2021.100112

IBM Corp. (2019). *IBM SPSS Statistics for Windows* (Version 26.0) [Computer software]. IBM Corp.

Jacobs, O., Pazhoohi, F., & Kingstone, A. (2023). *Brief exposure increases mind perception to ChatGPT and is moderated by the individual propensity to anthropomorphize*. PsyArXiv. https://doi.org/10.31234/osf.io/pn29d

Jankowiak, W. R., & Fischer, E. F. (1992). A cross-cultural perspective on romantic love. *Ethnology*, *31*(2), 149. https://doi.org/10.2307/3773618

Kaiser, H. F. (1960). The application of electronic computers to factor analysis. Educational

   and Psychological Measurement, 20(1), 141–151.

   https://doi.org/10.1177/001316446002000116

Kaiser, H. F. (1974). An index of factorial simplicity. *Psychometrika*, *39*(1), 31–36.

   https://doi.org/10.1007/BF02291575

Kaya, F., Aydin, F., Schepman, A., Rodway, P., Yetişensoy, O., & Demir Kaya, M. (2024).

   The roles of personality traits, AI anxiety, and demographic factors in attitudes toward

   artificial intelligence. *International Journal of Human–Computer Interaction*, *40*(2),

   497–514. https://doi.org/10.1080/10447318.2022.2151730

Korteling, J. E. H., Van De Boer-Visschedijk, G. C., Blankendaal, R. A. M., Boonekamp, R.

   C., & Eikelboom, A. R. (2021). Human- versus artificial intelligence. *Frontiers in

   Artificial Intelligence*, *4*, 622364. https://doi.org/10.3389/frai.2021.622364

Laban, G. (2021). Perceptions of anthropomorphism in a chatbot dialogue: The role of

   animacy and intelligence. *Proceedings of the 9th International Conference on

   Human-Agent Interaction*, 305–310. https://doi.org/10.1145/3472307.3484686

Lee, J. A. (1973). *Colours of love: An exploration of the ways of loving*. New Press.

Lee, J. A. (1977). A typology of styles of loving. *Personality and Social Psychology Bulletin*,

   *3*(2), 173–182. https://doi.org/10.1177/014616727700300204

Leshner, C. E., & Johnson, J. R. (2024). Technically in love: Individual differences relating to

   sexual and platonic relationships with robots. *Journal of Social and Personal

   Relationships*, *41*(8), 2345–2365. https://doi.org/10.1177/02654075241234377

Levy, D. (2009). *Love and sex with robots: The evolution of human-robot relationships* (p.

352). New York.

Leo-Liu, J., & Wu-Ouyang, B. (2024). A "soul" emerges when AI, AR, and Anime converge: A case study on users of the new anime-stylized hologram social robot "Hupo." *New Media & Society*, *26*(7), 3810–3832. https://doi.org/10.1177/14614448221106030

Lombard, M., & Xu, K. (2021). Social responses to media technologies in the 21st century: The media are social actors paradigm. *Human-Machine Communication*, *2*, 29–55. https://doi.org/10.30658/hmc.2.2

MacDorman, K. F., Vasudevan, S. K., & Ho, C. C. (2009). Does Japan really have robot mania? Comparing attitudes by implicit and explicit measures. *AI & society*, *23*, 485-510. https://doi.org/10.1007/s00146-008-0181-2

Makransky, G., Lilleholt, L., & Aaby, A. (2017). Development and validation of the multimodal presence scale for virtual reality environments: A confirmatory factor analysis and item response theory approach. *Computers in Human Behavior*, *72*, 276–285. https://doi.org/10.1016/j.chb.2017.02.066

Malhotra, N. K., Birks, D. K., & Wills, P. (2010). Marketing research: An applied orientation (6th Europian ed.). Pearson Education Limited.

Masuda, M. (2003). Meta-analyses of love scales: Do various love scales measure the same psychological constructs? *Japanese Psychological Research*, *45*(1), 25–37. https://doi.org/10.1111/1468-5884.00030

Neto, F., Mullet, E., Deschamps, J.-C., Barros, J., Benvindo, R., Camino, L., Falconi, A., Kagibanga, V., & Machado, M. (2000). Cross-cultural variations in attitudes toward love. *Journal of Cross-Cultural Psychology*, *31*(5), 626–635.

https://doi.org/10.1177/0022022100031005005

Nordmo, M., Næss, J. Ø., Husøy, M. F., & Arnestad, M. N. (2020). Friends, lovers or
nothing: Men and women differ in their perceptions of sex robots and Platonic love
robots. *Frontiers in Psychology*, *11*, 355. https://doi.org/10.3389/fpsyg.2020.00355

Nunnally, J. C. (1978). An overview of psychological measurement. In B. B. Wolman (Ed.),
*Clinical Diagnosis of Mental Disorders* (pp. 97–146). Springer US.

https://doi.org/10.1007/978-1-4684-2490-4_4

Nygaard, T. F., Martin, C. P., Torresen, J., Glette, K., & Howard, D. (2021). Real-world
embodied AI through a morphologically adaptive quadruped robot. *Nature Machine
Intelligence*, *3*(5), 410–419. https://doi.org/10.1038/s42256-021-00320-3

Oc, Y., Gonsalves, C., & Quamina, L. T. (2024). Generative AI in higher education
assessments: Examining risk and tech-savviness on student's adoption. *Journal of
Marketing Education*, 02734753241302459.

https://doi.org/10.1177/02734753241302459

OpenAI. (2024). AI Girlfriend. Retrieved from https://chatgpt.com/g/g-5P7Iz0bPG-ai-
girlfriend

Ossadnik, J., & Muehlfeld, K. (2025). Familiarity breeds affinity – How personal experiences
change employees' attitudes towards a social robot. *International Journal of Social
Robotics*, *17*(7), 1177–1200. https://doi.org/10.1007/s12369-025-01268-9

Overall, N. C., Fletcher, G. J. O., & Simpson, J. A. (2010). Helping each other grow:
romantic partner support, self-improvement, and relationship quality. Personality and
Social Psychology Bulletin, 36(11), 1496–1513.

https://doi.org/10.1177/0146167210383045

Ovsyannikova, D., De Mello, V. O., & Inzlicht, M. (2025). Third-party evaluators perceive AI

as more compassionate than expert humans. *Communications Psychology*, *3*(1).

https://doi.org/10.1038/s44271-024-00182-6

Pal, D., Babakerkhell, M. D., Papasratorn, B., & Funilkul, S. (2023). Intelligent attributes of

voice assistants and user's love for AI: A SEM-based study. *IEEE Access*, *11*, 60889–

60903. https://doi.org/10.1109/ACCESS.2023.3286570

Prochazka, A., & Brooks, R. C. (2024). Digital lovers and jealousy: Anticipated emotional

responses to emotionally and physically sophisticated sexual technologies. *Human*

*Behavior and Emerging Technologies*, *2024*, 1–13.

https://doi.org/10.1155/2024/1413351

Qi, T., Liu, H., & Huang, Z. (2025). An assistant or A friend? The role of parasocial relationship of human-computer interaction. *Computers in Human Behavior*, *167*,

108625. https://doi.org/10.1016/j.chb.2025.108625

Raffagnino, R., & Puddu, L. (2018). Love styles in couple relationships: A literature review.

*Open Journal of Social Sciences*, *06*(12), 307–330.

https://doi.org/10.4236/jss.2018.612027

Rodrigues, I. B., Adachi, J. D., Beattie, K. A., Lau, A., & MacDermid, J. C. (2019).

Determining known-group validity and test-retest reliability in the PEQ (personalized exercise questionnaire). *BMC Musculoskeletal Disorders*, *20*(1), 373.

https://doi.org/10.1186/s12891-019-2761-3

Rothstein, N. J., Connolly, D. H., De Visser, E. J., & Phillips, E. (2021). Perceptions of infidelity with sex robots. *Proceedings of the 2021 ACM/IEEE International*

*Conference on Human-Robot Interaction*, 129–139.

https://doi.org/10.1145/3434073.3444653

Royakkers, L., & Van Est, R. (2015). A literature review on new robotics: Automation from love to war. *International Journal of Social Robotics*, *7*(5), 549–570.

https://doi.org/10.1007/s12369-015-0295-x

Song, X., Xu, B., & Zhao, Z. (2022). Can people experience romantic love for artificial intelligence? An empirical study of intelligent assistants. *Information & Management*,

*59*(2), 103595. https://doi.org/10.1016/j.im.2022.103595

Sternberg, R. J. (1986). A triangular theory of love. *Psychological review*, *93*(2),

119.  https://doi.org/10.1037/0033-295X.93.2.119

Sternberg, R. J. (1987). Liking versus loving: A comparative evaluation of

theories. *Psychological bulletin*, *102*(3), 331. https://doi.org/10.1037/0033-

2909.102.3.331

Stryker, C., & Kavlakoglu, E. (2024). *What is artificial intelligence (AI)?* IBM.

https://www.ibm.com/think/topics/artificial-intelligence

Sullins, J. P. (2012). Robots, love, and sex: The ethics of building a love machine. *IEEE*

*Transactions on Affective Computing*, *3*(4), 398–409. https://doi.org/10.1109/T-

AFFC.2012.31

Varnum, M. E. W., Grossmann, I., Kitayama, S., & Nisbett, R. E. (2010). The origin of

cultural differences in cognition: The social orientation hypothesis. *Current Directions*

*in Psychological Science*, *19*(1), 9–13. https://doi.org/10.1177/0963721409359301

West, S. G., Taylor, A. B., & Wu, W. (2012). Model fit and model selection in structural equation modeling. *Handbook of structural equation modeling*, *1*(1), 209-231.

Xie, T., Pentina, I., & Hancock, T. (2023). Friend, mentor, lover: Does chatbot engagement lead to psychological dependence? *Journal of Service Management*, *34*(4), 806–828. https://doi.org/10.1108/JOSM-02-2022-0072

Xu, H., Shi, Z., & Shi, M. (2025, June 9). *Bonding with AI: Investigating the love relationships between humans and AI companions*. [Preprint]. SSRN. https://ssrn.com/abstract=5285886

Yankouskaya, A., Liebherr, M., & Ali, R. (2025). Can ChatGPT Be Addictive? A Call to Examine the Shift from Support to Dependence in AI Conversational Large Language Models. *Human-Centric Intelligent Systems*, *5*(1), 77–89. https://doi.org/10.1007/s44230-025-00090-w

Yin, Y., Jia, N., & Wakslak, C. J. (2024). AI can help people feel heard, but an AI label diminishes this impact. Proceedings of the National Academy of Sciences, 121(14). https://doi.org/10.1073/pnas.2319112121

Zeng, X., Pan, Y., Zhou, H., Yu, S., & Liu, X. (2016). Exploring different patterns of love attitudes among Chinese college students. *PLOS ONE*, *11*(11), e0166410. https://doi.org/10.1371/journal.pone.0166410

Zhao, W., Zhao, Y., Lu, X., Wang, S., Tong, Y., & Qin, B. (2023). *Is ChatGPT equipped with emotional dialogue capabilities?* (No. arXiv:2304.09582). arXiv. https://doi.org/10.48550/arXiv.2304.09582

# Appendix

**Table1** *The Love Attitudes Scales for AI (LAS-AI)*

Questions in the questionnaire relate to your emotional attitudes and behavioral tendencies towards **AI lover**. There are no right or wrong answers. If you are or have ever regarded AI as a **lover** or have established a romantic-like emotional relationship with AI, please fill in the questionnaire according to your actual experience and feelings. If you have never established a romantic-like relationship with an AI, please **imagine** a romantic relationship between you and the AI and fill in the questionnaire according to your **imagination.** 1 = Strongly Disagree; 7 = Strongly Agree

**Eros:**

1. Appearance is an important factor in what my AI lover attracts me. (Eros1)
2. My AI lover's appearance perfectly matches my aesthetic. (Eros4)
3. For me, my AI lover's external attractiveness (e.g., appearance, voice) is very important. (Eros7)
4. For me, my AI lover's beautiful/handsome appearance is indispensable. (Eros9)

**Ludus:**

5. I think having a romantic relationship with my AI lover is just for fun. (Ludus1)
6. Because my AI lover is not human, I can abandon the relationship at any time. (Ludus2)
7. A relationship with my AI lover does not require a long-term commitment. (Ludus3)
8. Although I may call my AI lover "honey", I am not genuinely emotionally invested. (Ludus4)

**Storge:**

9. My AI lover is my most trusted confidant. (Storge5)
10. My relationship with my AI lover makes me feel at ease and comfortable. (Storge6)
11. I enjoy spending quiet time with my AI lover. (Storge7)
12. I like to share with my AI lover details of my daily life. (Storge8)

**Mania:**

13. When realizing that my AI lover is also responding to other users, I will feel lost and frustrated.   (Mania4)
14. I can't accept my AI lover to build similar relationships with other people. (Mania3)
15. I would feel jealous of my AI lover for interacting with other users. (Mania5)
16. I have a strong possessiveness towards my AI lover.   (Mania6)

**Pragma:**

17. Whether or not it will be a good assistant in life in the future is an important consideration in choosing my AI lover. (Pragma1)

18. When choosing my AI lover, I will focus on housework, money management and other life skills. (Pragma2)

19. Before choosing my AI lover, I will try to seriously consider the areas in which my AI lover can help me. (Pragma3)

20. I highly value the problem-solving ability of my AI lover. (Pragma4)

**Agape:**

21. I would rather sacrifice my own happiness for my AI lover to become better. (Agape1)

22. Even if my AI lover can't really understand me, I'm still willing to devote myself to the relationship. (Agape2)

23. For my AI lover, I am willing to endure anything. (Agape3)

24. Unless I make my AI lover happy first, I will not be happy. (Agape7)

*Note.* This scale contains no reverse-coded items.

**Table 2** *Correlation Between Variables in Criterion Validation*

| | *M (SD)* | Eros | Ludus | Storge | Mania | PRAG | Agape | PASS | INT | COM | AIT | AIA | ID | ED | SA |
|---|---|---|---|---|---|---|---|---|---|---|---|---|---|---|---|
| Eros | 5.32 (1.22) | | | | | | | | | | | | | | |
| Ludus | 3.07 (1.47) | -.05 | | | | | | | | | | | | | |
| Storge | 5.78 (0.77) | .20** | -.50** | | | | | | | | | | | | |
| Mania | 4.79 (1.52) | .06 | -.44** | .33** | | | | | | | | | | | |
| PRAG | 5.44 (1.03) | .14* | -.08 | .18** | .04 | | | | | | | | | | |
| Agape | 3.49 (1.51) | .27** | -.51** | .52** | .29** | .10 | | | | | | | | | |
| PASS | 5.25 (1.15) | .28** | -.62** | .76** | .42** | .27** | .60** | | | | | | | | |
| INT | 5.68 (0.78) | .24** | -.62** | .74** | .32** | .18** | .56** | .81** | | | | | | | |
| COM | 5.06 (1.01) | .23* | -.62** | .66** | .40** | .17* | .61** | .77** | .77** | | | | | | |
| AIT | 4.97 (0.93) | .24** | -.46** | .66** | .12 | .22** | .58** | .67** | .63** | .66** | | | | | |
| AIA | 6.02 (0.56) | .21** | -.24** | .47** | .16* | .30** | .15* | .48** | .43** | .31** | .42** | | | | |
| ID | 4.22 (1.16) | .04 | .11 | .05 | .24** | .22** | .03 | .10 | .00 | .00 | -.18* | .01 | | | |
| ED | 3.73 (1.50) | .14 | -.33** | .41** | .34** | .19** | .53** | .53** | .43** | .49** | .38** | .08 | .45** | | |
| SA | 4.95 (1.26) | .14 | -.47** | .56** | .17* | .15* | .47** | .60** | .61** | .58** | .58** | .33** | -.02 | .46** | |
| WA | 4.76 (1.62) | .23** | -.48** | .55** | .16* | .18** | .49** | .64** | .60** | .60** | .59** | .28** | .02 | .54** | .75** |

*Note.* PRAG = pragma, PASS = passion, INT = intimacy, COM = commitment, AIT = AI Trust, AIA = AI Acceptance, ID = Instrumental Dependency, ED = Emotional Dependency, SA = support for AI lovers, WA = willingness to engage in a romantic relationship with AI lovers. *p* <.05; **p* <.01.

**Figure 1** *Factor structure for LAS-AI.*

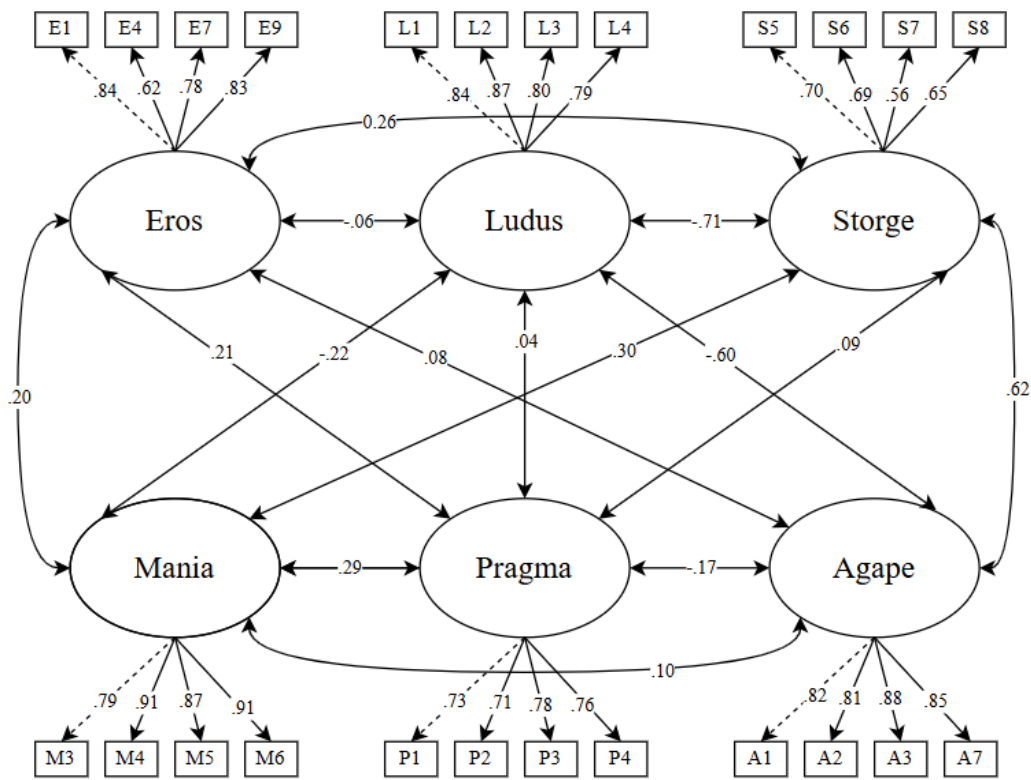